\documentclass{emulateapj}
\usepackage{times}

\def\la{\mathrel{\mathpalette\fun <}}
\def\ga{\mathrel{\mathpalette\fun >}}

\def\simpropto{\lower.2ex\hbox{$\; \buildrel \sim \over \propto \;$}}

\def\fun#1#2{\lower0.837ex\vbox{\baselineskip0ex\lineskip0.209ex
  \ialign{$\mathsurround=0ex#1\hfil##\hfil$\crcr#2\crcr\sim\crcr}}}

\def\msun{M_\odot}

\def\sles{\lower2pt\hbox{$\buildrel {\scriptstyle <}
   \over {\scriptstyle\sim}$}}

\def\sgreat{\lower2pt\hbox{$\buildrel {\scriptstyle >}
   \over {\scriptstyle\sim}$}}

\def\la{\mathrel{\mathpalette\fun <}}
\def\ga{\mathrel{\mathpalette\fun >}}

\def\msun{M_\odot}

\begin{document}

\title{GRB 110328A/{\it Swift} J164449.3+573451:
  The Tidal Obliteration of a Deeply Plunging Star?}
\shortauthors{CANNIZZO, TROJA, \& LODATO}
\author{
 J.~K.~Cannizzo\altaffilmark{1,2},
 E.~Troja\altaffilmark{3,4}, 
 and
 G. Lodato\altaffilmark{5}}

\altaffiltext{1}{CRESST and Astroparticle Physics Laboratory NASA/GSFC, Greenbelt, 
                MD 20771, USA; 
               John.K.Cannizzo@nasa.gov}
\altaffiltext{2}{Department of Physics, University of Maryland, Baltimore County, 
                 1000 Hilltop Circle, Baltimore, MD 21250, USA}
\altaffiltext{3}{Astroparticle Physics Laboratory NASA/GSFC,
                 Greenbelt, MD 20771, USA}
\altaffiltext{4}{NASA Postdoctoral Program}
\altaffiltext{5}{Universit\`a degli Studi di Milano, Via Celoria 16, Milano, Italy}

\begin{abstract}
We examine the tidal disruption
  event scenario   
   to explain Sw 1644+57,
  a powerful 
  and persistent
  X-ray source which suddenly became active
as GRB 110328A.
%
    The precise localization at the center of a $z = 0.35$ galaxy
  argue for activity of the central engine as the underlying cause.
 We look at the
   suggestion by Bloom et al of
  the
    possibility of a  tidal disruption
  event (TDE).
   We argue that Sw 1644+57 cannot be explained
  by the traditional TDE model in which
  the periastron distance is close to the tidal disruption
   radius $-$ 
   three independent lines of argument
  indicate the orbit must be deeply plunging or 
else the powerful jet we are observing could not be
produced.   
 These arguments stem from   
 (i) comparing the early X-ray light 
  curve to the expected theoretical fallback rate,
  (ii) looking at the time of transition
   to disk-dominated decay,
  and (iii)
   considering the TDE rate.
         Due to the extreme excess in the tidal force
above that which would be required  minimally to disrupt
  the star in a deeply plunging orbit
   at periastron,
    we suggest this scenario 
    might be referred
    to more descriptively as 
    a TOE (tidal obliteration event)
   rather than a TDE.
\end{abstract}


\section{Introduction}

Given the massive black holes (MBHs) known to 
reside at the centers of most galaxies
  (Magorrian et al. 1998), 
   an inevitable occurrence will be the
   occasional
errant stellar wandering  into the inner
   galactic environment, where
the prospect for a tidal disruption
  and subsequent swallowing of
  some fraction of the ill-fated star may ensue
  (Rees 1988, 1990).
   A complete study of the tidal disruption process
  was carried out recently by Strubbe \& Quataert (2009, 
  hereafter SQ09) and Lodato \& Rossi (2011, hereafter LR11).
  For MBHs with greater than $10^8\msun$,
 the tidal disruption radius lies inside
the event horizon for a typical main sequence
  star, so the star would be swallowed whole.
   The rate of such events is calculated to be
  $\sim$10$^{-5}-10^{-3}$ yr$^{-1}$ galaxy$^{-1}$,
depending on galaxy type (Wang \& Merritt 2004).
 Wang \& Merritt 
derive a rate  $\sim$10$^{-5}$ yr$^{-1}$ Mpc$^{-3}$,
  which, given a local galaxy space density of 
 $\sim$0.01        Mpc$^{-3}$ (Driver et al. 2005),
  yields an effective rate per galaxy of
  about  $10^{-3}$ yr$^{-1}$.
  This  is
 $\sim$10 times higher than
  previous estimates (e.g., Cohn \& Kulsrud 1978).
%
   The best candidates to date 
   are presented in 
 van Velzen et al. (2010) $-$  
two strong examples of optical flares
from tidal disruption
events in archival SDSS data,
  from which the authors determine
   a rate of
   tidal disruption 
events (TDEs) 
 $\sim$3$^{+4}_{-2} \times 10^{-5}
 $ yr$^{-1}$  galaxy$^{-1}$, about a factor
   of 30 below the Wang \& Merritt
  theoretical estimate.
    There were also
  earlier claims of TDEs
(e.g., 
  Komossa et al. 2004, 2008;
   Gezari et al. 2008, 2009).
%
%


After the star comes undone, roughly half
 of the remnants remains on bound,
  highly eccentric elliptical orbits
                     and accretes;
   the other half lies on unbound, hyperbolic
   orbits and is ejected (Lacy,
  Townes, \& Hollenbach 1982, hereafter LTH82,
   Rees 1988, 1990).
Theoretical research on accretion of tidal
debris from tidal disruptions
has traditionally been divided into two phases:
  (1) the immediate unbinding of the star and accretion
  of streamers of gas, for which estimates a fallback
  rate ${\dot M}\simpropto t^{-5/3}$ 
    (Rees 1988; Evans \& Kochanek 1989;
   Lodato, King, \& Pringle 2009, SQ09,
  Ramirez-Ruiz \& Rosswog 2009,
  Guillochon et al. 2009,
     LR11),
and (2) the subsequent disk accretion, for which the freely
expanding outer disk enforces 
   ${\dot M}\simpropto t^{-19/16}$
   for a thin disk 
     (Cannizzo, Lee, \& Goodman 1990, hereafter CLG90,
  Ulmer 1999, SQ09).
 and   ${\dot M}\simpropto t^{-4/3}$
   for a thick disk 
     (Cannizzo \& Gehrels 2009, hereafter CG09).
  The longer term evolution,  
   which  can be cast in terms
 of accretion disk physics, predicts 
  the tidal disruption flare to be  a
  strong UV source, $\sim$10$^{43}$ erg s$^{-1}$,
   which a duration of months to years.
  These expectations are nominally satisfied
in the handful of secure candidates (van Velzen et al. 2010).

       Sw 1644+57
  was triggered by the {\it Swift}/BAT as a low
SNR image  trigger requiring a 2000s integration (Cummings et al. 2011;
  see Burrows et al. 2011 for full details of
the {\it Swift} analysis).
%
%
  Ground-based NIR and radio observations provided the first
clear evidence connecting the X-ray/gamma-ray source
with the galaxy nucleus (Berger et al. 2011).
Precise localizations by {\it HST} 
  (Fruchter et al. 2011) and {\it Chandra} (Levan
et al. 2011a) confirmed the transient 
   to lie at the center of its galaxy.
%
   Sw 1644+57   is
  at a redshift $z=0.3534\pm0.0002$ (Levan et al. 2011ab).
 The inferred R-band magnitude at that distance
  implies $M_R \simeq -18$ (Fruchter et al. 2011),
   which implies a $\sim$10$^7\msun$
  black hole at the galactic center, if the
galaxy follows the Magorrian relation (Magorrian et al. 1998).
  Burrows et al. (2011) derive a central BH mass $2\times 10^7\msun$
based on the Magorrian relation, and a lower limit$\sim$$7\times 10^6\msun$
  based on X-ray variability.
%
Quataert \& Kasen (2011) point out however
  that the signal-to-noise of
      the XRT data  is not sufficient to constrain 
  significant variability on $<10$ s timescales,
  therefore the central BH mass could be much lower.
  Similarly, 
  the host galaxy brightness places only an upper limit on 
  the BH mass from the
      Magorrian relation, since no ``bulge''
     is actually resolved. 
  In addition,
  Miller \& G\"ultekin (2011) use
   empirical relations relating
  central black hole mass
   to radio and X-ray luminosity 
  for Sw1644+57 to argue
   for a central 
mass $\log (M_{\rm MBH}/\msun) = 5.9 \pm 1.1$.
   In view of these considerations
  and the uncertainty associated with applying the
   Magorrian relation, in this work
    we seriously entertain
  the possibility that
   $M_{\rm MBH}$ may lie between $\sim10^6\msun$ and $\sim10^7\msun$.
 
   The positional coincidence gives strong   credence
  to the notion of an event associated with the putative
    MBH at the galaxy center,
    quite probably a tidal disruption event
   (Bloom et al. 2011ab,    
  Shao et al. 2011),
  and yet the contrast between theoretical expectation
(UV source at $10^{43}$ erg s$^{-1}$)
and observation 
 (gamma-ray/X-ray source at 
    $10^{47}-10^{48}$ erg s$^{-1}$)
   lead to the unavoidable conclusion that we 
   are viewing a strongly beamed event.
   In fact, the jet that we must be observing
  is completely  divorced from the physical properties
 of the fallback disk, other than relying on accretion
 as a  power supply.
 

In this work we estimate the
   rate of accretion within a transient disk formed from the 
tidal debris of a disrupted star,
   and estimate  a temperature in the inner disk.
  We also calculate the expected rates, both
   for {\it Swift}/BAT  and  all-sky X-ray monitors
  with $\sim$mCrab sensitivity.


\section{General Framework}

\subsection{Accretion Properties}

   Given the existence of a compact mass $M_{\rm MBH}$
  and the sudden introduction of a much smaller mass
   of stellar debris
 $\Delta M$  at $\sim 1-10 R_S$,
  where the Schwarzschild radius 
 $R_S = 2GM_{\rm MBH}/c^2$,
  what will be the
  subsequent rate of accretion
  onto the MBH as a function of time?

In simple physical terms,
the disruption of the star occurs when the
  star comes closer to the MBH
     than a tidal disruption radius $R_T$
  which is determined by 
   demanding that the  mean density
  of the volume of space
  enclosed by $R_T$, i.e., 
   $M_{\rm MBH}/(4\pi R_T^3/3)$,
 be  equal to the density of the star.
   This gives
\begin{equation}
R_T \simeq R_*(M_{\rm MBH}/M_*)^{1/3} = 1.50\times 10^{13} \ {\rm cm}
         \ m_*^{-1/3} 
          r_* 
         m_{b7}^{1/3},
\end{equation}
  where
    $m_*=M_*/\msun$, 
    $r_*=R_*/R_{\odot}$,
    and
  $m_{b7}=M_{\rm MBH}/10^7\msun$
     (Rees 1988, 1990).
  Expressing $R_T$ in terms of $R_S=2.95\times10^{12}$ cm $m_{b7}$
    gives
\begin{equation}
{R_T\over R_S} = 5.1  
         \ m_*^{-1/3} 
          r_* 
         m_{b7}^{-2/3},
\end{equation} 
   Setting this equal to unity and solving for $m_{b7}$,
we see that for $m_{b7}\ga 10$, the tidal radius will lie inside  
 the Schwarzschild radius, and a main sequence star will be swallowed whole.
  For a maximal Kerr BH in which $R_S=GM_{\rm MBH}/c^2$, there
   would still be another factor $2^{3/2}$
   increase in $M_{\rm MBH}$
    before this limit is  reached.
 An equivalent way of visualizing 
this is in terms of the mean density 
enclosed by $R_S$.
    For a nonspinning (i.e., Schwarzschild) BH,
the mean density out to $R_S$ is $\rho_{\rm MBH}=185$ g cm$^{-3}$
   $m_{b7}^{-2}$. For comparison $\rho_{\odot}  = 1.4$ g cm$^{-3}$.
    Stellar disruption occurs if $\rho_{\rm MBH} > \rho_*$
and $R_S < R_P < R_T$,
   where $R_P$ is the periastron 
   distance of the star.
 
  We define $\xi\equiv R_P/R_T$. 
The requirement that the TDE occur
   outside the event horizon
  enforces
%
\begin{equation}
\xi > \chi R_S/R_T \approx 
        0.2 \  m_*^{1/3} r_*^{-1} m_{b7}^{2/3} \chi,
\end{equation}
  where $\chi=1$ for a Schwarzschild BH and 
        $\chi=0.5$  for a Kerr BH.

 We now look at the long term evolution
of the resultant mass accretion rate onto the MBH, 
both in terms of stellar debris fallback 
and accretion disk.

\subsubsection{Debris from Stellar Fall-Back}

 Earlier studies give an expression, 
   derived in the
  Newtonian limit, for the time scale of 
return of the most bound stellar material to
periastron (LTH82, Rees 1988, cf. Equation [4] of LR11)
\begin{equation}
t_{\rm fallback} \simeq {\pi\over 2^{1/2}} 
  \left(R_P\over R_*\right)^{3/2}
     t_P,
\end{equation} 
where the dynamical time
   at periastron $t_P = (G M_{\rm MBH}/R_P^3)^{-1/2}$.
Scaling $R_P$ in terms of $R_T$ gives
\begin{equation}
t_{\rm fallback} = 130  \ {\rm d} \ 
   m_{b7}^{1/2} 
      \xi^3 m_*^{-1} r_*^{3/2}.
\end{equation} 
  The condition on $\xi$ from Equation (3) 
   gives the lower limit
%
\begin{equation}
t_{\rm fallback} > 0.98  \ {\rm d} \ 
   \chi^3
      m_{b7}^{5/2} 
      r_*^{-3/2}.
\end{equation}
The accretion rate due to fallback is then 
 (e.g., LTH82, Rees 1988, SQ09, LR11) 
\begin{equation}
{\dot M}_{\rm fallback} \simeq  {1\over 3}
           {M_* \over t_{\rm fallback}}
      \left( t
            \over t_{\rm fallback} \right)^{-5/3},
\end{equation} 
  so that ${\dot M}_{\rm fallback} 
   \propto m_{b7}^{1/3} \xi^2 $.
  The total fallback mass 
    $\Delta M(t_1, t_2)
   = \int_{t_1}^{t_2}  {\dot M}_{\rm fallback} dt
\rightarrow 0.5 M_*$
  in the limit $t_1\rightarrow t_{\rm fallback}$
and $t_2\rightarrow \infty$.
  Note that ${\dot M}_{\rm fallback}=0$
 for $t<  t_{\rm fallback}$, and we can 
 thus identify $t=t_{\rm fallback}$ as the time of the 
  earliest significant BAT activity, $\sim2$ d before the GRB 110328A trigger.

 We can estimate the basic dynamical parameters 
 of the tidal disruption event in the following way. 
  Superposed on the long-term 
  X-ray light curve of Sw 1644+57,
   Levan et al. (2011b)
     and
   Bloom et al. (2011b) 
  show  a $t^{-5/3}$
   decay law. This suggests 
 that the X-ray luminosity
   tracks the rate of return of the fallback material 
  to pericenter. 
  In this work however, we argue that only 
  the early light curve, $t\la6$ d,  is representative of
 the stellar debris fallback;  the subsequent evolution
  is dominated by the accretion disk.
As mentioned above, the very high X-ray 
 luminosity points to a strongly beamed source, with a 
  beaming factor $\epsilon_{\rm b}\sim 0.01$, so that the intrinsic 
 jet luminosity $L_{\rm j}$ is related to the observed $L_{\rm X}$ by
\begin{equation}
L_{\rm j} = \epsilon_{\rm b} L_{\rm X}. 
\end{equation}
 Let us assume that the jet luminosity is a fraction $f$ of
    the kinetic energy carried by the outflowing material:
\begin{equation}
L_{\rm j} = f \dot{M}_{\rm j}c^2, 
\end{equation}
where we have also assumed that the outflow velocity 
 $v\sim c$. The outflow rate itself is a fraction 
 $\epsilon_{\rm j}$ of the fallback rate $\dot{M}_{\rm fallback}$, 
 where, for a mildly super-Eddington accretion flow, we expect
  $\epsilon_{\rm j} \sim 0.1$ (SQ09, LR11, Dotan \& Shaviv 2011). 
      We do not wish to restrict ourselves to baryonic jets;
  this same analysis could be applied to a Poynting flux jet,
  with $f$ denoting a mass-energy equivalent flux.
 We can thus obtain an estimate of the fallback rate 
 $\dot{M}_{\rm fallback}$ as a function of the observed X-ray luminosity:
\begin{equation}
\dot{M}_{\rm fallback}(t) = \frac{\epsilon_b}{f\epsilon_j}\frac{L_{\rm X}(t)}{c^2}.
\end{equation}
Using Equations (5) and (7), we can thus obtain an estimate of the fallback time and of the penetration factor $\xi$:
\begin{equation}
t_{\rm fallback} = \left(\frac{\epsilon_b}{f\epsilon_j}\frac{3L_X(t)t}{M_*c^2}\right)^{3/2} t,
\end{equation}
\begin{equation}
\xi = \left(\frac{t}{130 \mbox{d}}\right)^{1/3}\left(\frac{\epsilon_b}{f\epsilon_j}\frac{3L_X(t)t}{M_*c^2}\right)^{1/2}m_*^{1/3}r_*^{-1/2}m_{b7}^{-1/6}.
\end{equation}
Note that $\xi$ is very weakly dependent on both 
the stellar mass and on the black hole 
 mass $\xi\propto (m_*m_{b7})^{-1/6}$ (Krolik \& Piran 2011).  
 We acknowledge the simplicity of the assumption that $f$ and and $\epsilon_j$ are 
   independent of the Eddington ratio of the accretion rate.
  There is observational evidence that AGN decrease in radio loudness 
  with decreasing Eddington ratio (e.g., Sikora et al. 2007).
   This may be the result of a fundamental change in the 
    underlying accretion disk structure, e.g., from a slim disk to a thin disk.
   Based on the total Sw1644+57 light curve to date, which has exhibited 
a  dynamic variation in averaged X-ray flux of $\la$2  orders of magnitude,
   our assumption 
    ${\dot f} = 
       {\dot {\epsilon_j}} = 
       {\dot {\epsilon_b}} = 
         0 $  
  may be at least crudely justified. 
   In other words,
   although  the accretion physics will likely change
 with Eddington ratio (i.e., slim-to-thin disk), this
  transition occurs at 
 a low Eddington ratio $\sim$0.01, 
   which is not applicable to date for Sw1644+57.

  Several works have 
 considered the observational 
  consequences of ultra-relativistic jets from TDEs prior
      to Sw 1644+57   (Farrar \& Gruzinov 2009, 
       Giannios \& Metzger 2011).
       Giannios \& Metzger find,  on the assumption of a baryonic
  jet, that  emission from the forward shock may be detectable
  for several years.  However, motivated 
 by  results from 
         recent groups studying 3D GRMHD jets  
  (Hawley \& Krolik 2006,
 McKinney \& Blandford 2009)
   which find a baryonic zone-of-exclusion in the jet beam, leading
to primarily Poynting flux jets, we do not restrict our focus
to  baryonic jets, and indeed we view  Poynting flux jets to
be more likely.
  On the other hand,
     the jet luminosity can be
             dominated by Poynting flux  even if  most of its rest
             mass is in baryons.
%
             If the jet remains Poynting dominated
out to the
   distance  where it begins to interact
             with the ISM, a forward shock with concomitant non-thermal emission
             can still be produced
    (e.g., Mimica, Giannios, \& Aloy 2009).
 A
             highly magnetized jet transfers energy
efficiently to the environment,
   and the
  Lorentz factor of the
            forward shock
   is related to the magnetization of the jet,
    not
 its
  Lorentz factor,
   which could be
  much less.

At $t=2.5$ d from the BAT trigger 
 the X-ray luminosity crosses $10^{47}$ erg s$^{-1}$. Inserting these values 
 into Equation (11),
   after taking into account the offset time 2 d
  to be added to the 2.5 d,
%
%
%
    we obtain $t_{\rm fallback} \simeq6480$ s 
       $(\epsilon_b/f \epsilon_j)^{3/2}
                                                \ll$ 2.5 d. 
 From Equation (12) we obtain an estimate of $\xi$:
\begin{equation}
\xi \simeq 0.0833 \left(\frac{\epsilon_{\rm b}}{0.01}\right)^{1/2} 
   \left(\frac{\epsilon_{\rm j}}{0.1}\right)^{-1/2}
     \left(f \over 0.1 \right)^{-1/2}(m_*m_{b7})^{-1/6}r_*^{-1/2}.
\end{equation} 
We thus see that the numbers involved point strongly 
 towards a deeply plunging event\footnote{Note that for a WD instead of a MS
  star, i.e., $R_*=5\times 10^8$ cm and $m_{b7}=10^{-3}$
   (Krolik \& Piran 2011), Equation (13) gives $\xi\simeq3.1$,
which would require an adjustment of the efficiency parameters
 to still yield a tidal disruption,  $\xi<1$.}
  with $\xi\ll 1$. 
If one assumes $\epsilon_b=0.01$ and $\epsilon_j=f=0.1$, 
  the resulting
 $\xi$ value would be smaller than the minimum given
  by  Equation
(6). This allows us to put reasonable bounds on
  the least constrained parameters, $f$ and $\epsilon_j$. 
 By requiring that $\xi$ be larger
than the minimum value, we get
\begin{equation}
\frac{f\epsilon_j}{\epsilon_b}< 0.18\chi^{-2/3}m_{b7}^{-5/3}r_*^{-1}.
\end{equation}
Unless ${f\epsilon_j}/{\epsilon_b}$ is much smaller than 
  the value above, the event must have been deeply plunging.
   Note that the
condition above is strongly dependent on the black hole mass.
   If the MBH mass were smaller, closer to $10^6M_{\odot}$, that would relax
the stringency of the deeply plunging constraint.

\subsubsection{Accretion Disk}

The need for a deeply plunging orbit can also
   be obtained considering the properties of the
   accretion disc formed by the debris. 
The process  of fallback does not directly supply
 gas into the MBH, but rather it
   returns material to $\sim$$R_P$
and therefore feeds an accretion disk.
   Given the small inferred 
   $t_{\rm fallback}$ value
   with respect to the total length of the X-ray light curve,
 it seems probable that at early times an accretion
  disk is established.
  An evaluation of $\int_{t_{\rm fallback}}^{t_2}
    {\dot M}_{\rm fallback} dt$
   using Equation (7) shows that 90\% of the 
  total available fallback mass $0.5 M_*$
  has already been supplied to the accretion disk
    by $t_2 = $32$t_{\rm fallback}$.
%
  It is therefore of interest to estimate the properties
 of the accretion disk, given a resident disk mass $\sim$$0.5\msun$.


One may estimate the long term accretion rate
in Sw 1644+57 by looking at the properties 
   of the accretion disk formed from the debris.
   We  idealize the disk
  as  extending from
the event horizon to  
       $\eta R_P$,
  where $\eta\simeq 2$. 
   For the high rates of accretion expected
 we use the super-Eddington
  slim disk considered by 
   CG09
  (see their Section 2.3.2)
  in which the radial distribution of surface density
 $\Sigma(R)$ in a roughly steady disk
      would be $\propto R^{-1/2}$,
 thus $\Sigma(r) = \Sigma_0 (R/R_P)^{-1/2}$.
Therefore integrating the mass distribution 
  $\int 2 \pi R dR \Sigma(R)$ out to 
$\eta R_P$, and setting it equal to the mass of the fallback
debris $\Delta M$, one obtains
\begin{equation}
 \Sigma_0 =  { \Delta M \over ( 2/3)\pi \eta^2 {R_P}^2 }
          =  {3\over { 8 \pi}}
       { \Delta M \over {(\eta/2)^2 \xi^2 {R_T}^2} }.
\end{equation}
   We  derive an estimate for the properties of the 
resultant disk using   algebraic scalings.
 Inverting Equation (24) of CG09 to obtain ${\dot M}_0$
    yields 
\begin{equation}
{\dot M}_0 = 1.63\times 10^{23} \ {\rm g} \ {\rm s}^{-1}
         \Sigma_0       m_{b7}^{ 1/2} r_{13}^{ 1/2} \alpha_{-1},
\end{equation}
   where the disk radius  $r_{13}=\eta R_P/10^{13}$ cm
  and
    $\alpha_{-1}$ is 
the Shakura \& Sunyaev (1973, hereafter SS73)
    $\alpha$ parameter in
   units of 0.1.
Substituting for $\Sigma_0$ gives
\begin{equation}
{\dot M}_0 = 7.72\times 10^{29} \ {\rm g} \ {\rm s}^{-1}
             \Delta M_{1/2}
          m_{b7}^{ 1/2} r_{13}^{-3/2}  
             \alpha_{-1},
\end{equation}
   where $\Delta M_{1/2}$ is the fallback debris mass
  in  units of $0.5\msun$.
Substituting
  for the disk radius $\eta R_P$
  in units of $10^{13}$ cm,  
       $r_{13}= 
  1.5 \eta \xi m_*^{-1/3} r_* m_{b7}^{1/3}$,
     gives
\begin{equation}
{\dot M}_0 = 4.21\times 10^{29} \ {\rm g} \ {\rm s}^{-1}
          (\eta/2)^{-3/2}   
          \xi^{-3/2}   
          m_*^{ 3/2} 
          r_*^{-3/2}   
         \alpha_{-1}.
\end{equation}
 Equation (18) only gives a measure of the initial value of the rate of
  accretion within the accretion disk formed rapidly by stellar fallback.
  The longer term evolution is given by CG09 
as
\begin{equation}
{\dot M}(t) = {\dot M}_0   \left( t
         \over t_0 \right)^{-4/3},
\end{equation}
  where 
\begin{eqnarray}
t_0&=  {  {r_P}^{3/2}   \over { 9/4 \  \alpha (GM_{\rm MBH})^{1/2}  }}  = {4\over 9}  \xi^{3/2} \   t_T  \   {\alpha}^{-1}    \nonumber \\
   & = 224 \ {\rm s}  \left(\xi\over 0.1\right)^{3/2}   
 m_*^{-1/2} r_*^{3/2}  \   \alpha_{-1}^{-1}, 
\end{eqnarray}
   and $t_T = (GM_{\rm MBH}/R_T^3)^{-1/2}$.
   The time $t_0$ is basically the viscous time for a
  thick disk at $R_P$.
%

    For a disk mass which varies as $t^{-\beta}$,
   the global
   viscous time scale $t_{\rm visc}(t) \equiv |{M_{\rm disk}}/{{\dot M}_{\rm disk}}|
 = \beta^{-1}t$ increases with time. The outer edge of an ``external'' accretion disk,
  defined by the lack of a confining torque as from a companion
  in an interacting binary, moves outward with time.
  The  fact that  $t_{\rm visc}(t) \propto t$ results from the spreading of
  the outer disk edge,
 and the viscous time is determined 
     by the slowest time scale in the system,
i.e., that at large radii.
  On the other hand, in comparing the time scales for mass
   fallback with a viscous time scale, the more relevant time
 within the disk is not that at the outer edge, but rather the response
  time of the disk at the radii where the fallback mass is deposited $t_0$.

\begin{figure}
\begin{center}
\includegraphics[height=125mm]{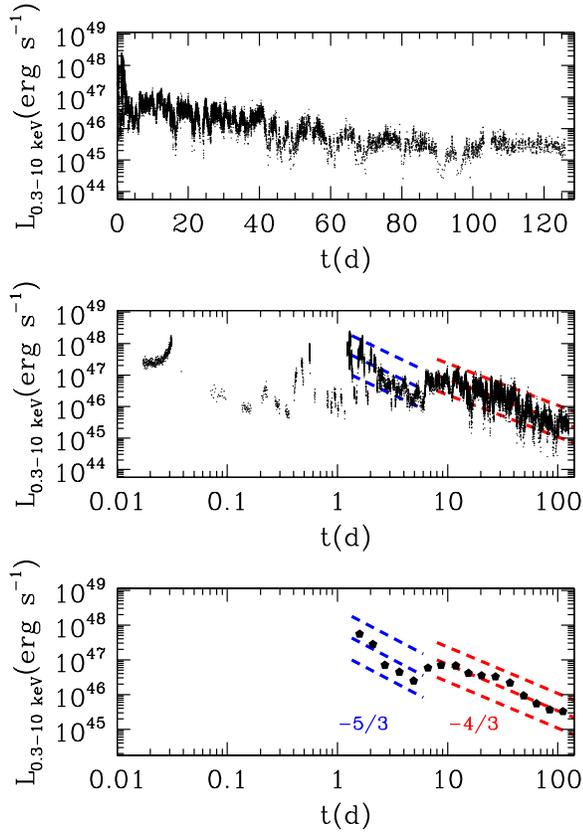}
\end{center}
\vskip -0.75cm
\caption{The 
  {\it Swift}/XRT light curve
  (Evans et al. 2007, 2009) for Sw 1644+57,
  plotted on both a log-linear (first panel) and log-log scale
  (second and third panels).
   In the third panel we show the 
  block-averages  of the
 the X-ray flux,  
    binned  in
  time in 0.125 dex bins.
  The blue lines indicate the putative stellar debris
   fallback slope $-5/3$ and the red lines show the fallback
  decline rate expected due to a super-Eddington slim disk, $-4/3$.
   The data do not strongly support either decay law in detail, 
    but are suggestive.
}
\end{figure}

\smallskip

  Figure 1 shows the inferred long-term X-ray luminosity
  for Sw 1644+57,
  taking into account
  the distance given by the redshift.
   The 
    X-ray flux 
      $\sim$10$^{47}-10^{48}$ erg s$^{-1}$
   is super-Eddington
    for $m_{b7}=0.1-1$, and
    also quite different spectrally
   from theoretical expectations if we
  were directly observing the disk.
   On the lower panels we indicate the expected slopes for
   stellar debris fallback and accretion due to the external
   disk formed by the stellar debris. 
   There appears to be a difference in the decay
   characteristics 
   at $t\simeq 6$d, which we interpret as the
   start of the disk-dominated decay.

   Figure 2 shows a comparison between the stellar debris fallback
rate  from Equation (7) and the disk accretion
   rate  from Equation (19) for $\xi=0.1$ and 0.2.
   For $m_{b7}=1$
    the transition point for the $\xi=0.1$ lines $t_{\times}=4.6$ d
  is close to the time $t\simeq6$ d in the Sw 1644+57 light curve
    where there appears to be a change in the decay characteristics.
   If this is indeed associated with the debris fallback/disk decay
  transition, it provides further evidence for a small   periastron
value $\xi\simeq 0.1$.
   For $m_{b7}=0.1$ the putative $t\simeq6$ d transition
   corresponds to  a larger  $\xi\simeq 0.18$.
   Given the results of Miller \& G\"ultekin (2011)  the latter
   value may be better motivated.
%
%
 We stress, however,
      the uncertainties
  associated with $m_{b7}$ and $\eta$, 
  and therefore our inferred $\xi$;
      we certainly do not claim a precise determination
   of $\xi$.  On the other hand, it is noteworthy
   that the uncertainties in 
                $f$, $\epsilon_b$, and $\epsilon_j$
    from the previous section do not enter into the $\xi_{\times}(t_{\times})$
   line of argument in this section, 
 and yet the inferred $\xi$ value
    obtained by adopting reasonable values for 
                $f$, $\epsilon_b$, and $\epsilon_j$
  and fitting the fallback decay to the early light curve
   are in line with that obtained using the putative $t_{\times}\simeq6$ d
  value in the light curve in conjunction with Equations (7) and (19).
%

Setting Equations (7) and (19) equal and solving for the time $t_{\times}$
  defining the transition point yields
\begin{equation}
t_{\times} = 1.25 \times 10^{10} \ {\rm s}
      \ m_*^{-3/2} 
      \ r_*^{3/2}
      \ m_{b7} 
      \ \alpha_{-1}
      \ \xi^{9/2}
      \ (\eta/2)^{9/2}.
\end{equation}
  For our favored parameters
   $m_*=r_*=\alpha_{-1}=1$,
  the transition point in the light curve
  to a disk dominated decay
  would lie at
\begin{equation}
\xi_{\times} = 0.106
  \left( t_{\times} \over
    { 6 \ {\rm d} } \right)^{2/9}
        m_{b7}^{-2/9}
     \left( \eta \over 2 \right)^{-1}.
\end{equation}

\begin{figure}
\begin{center}
\includegraphics[height=125mm]{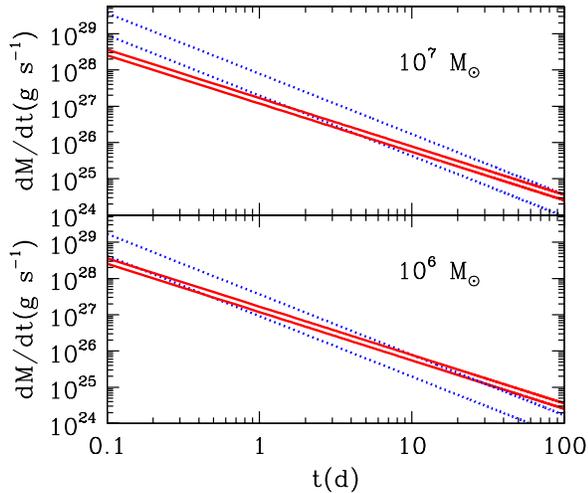}
\end{center}
\vskip -4.95cm
\caption{The theoretical accretion rates 
    for stellar debris fallback (blue) and
  a super-Eddington slim disk (red)
    for the parameters given in the text.
   The two sets of lines indicate $\xi=0.1$ and 0.2.
  In the upper panel we take $m_{b7}=1$ and 
  in the lower panel 
  $m_{b7}=0.1$. 
             For $m_{b7}=1$ the transitions
lie   at $t=4.59$ d ($\xi=0.1$, $t_{\rm fallback}=3.11$ hr) and
         $t=103$ d ($\xi=0.2$, $t_{\rm fallback}=1.04$ d), with $\xi_{\times} = 0.106$
   corresponding to $t_{\times} =6$ d;
            for $m_{b7}=0.1$ the transitions
  are at $t=0.46$ d ($\xi=0.1$, $t_{\rm fallback}=3540$ s) and
         $t=10.3$ d ($\xi=0.2$, $t_{\rm fallback}=7.87$ hr), with $\xi_{\times} = 0.177$
   corresponding to $t_{\times} =6$ d.
  }
\end{figure}

\subsection{Event Rate}

Why have such events not been seen previously?
  As noted earlier, Sw 1644+57 was nearly missed due
to its faintness in the {\it Swift}/BAT,
  which is a $\sim$mCrab instrument.
 Its brightness in the {\it Swift}/XRT
arises 
   because of the XRT's $\sim$$\mu$Crab sensitivity.
 
{\bf {\it Swift}/BAT rate:}
       The highest SNR flare
 from Sw 1644+57
     seen in BAT could
 have been detected out to $z_{\rm max}\simeq0.8$
\footnote{
  The redshift $z_{\rm det}=0.35$
   for Sw J1644 
  corresponds   to
only $\sim$12\%
of the estimated sensitivity 
  volume. At present, based on a single 
detection,
it is still premature to speculate on a possible cosmic evolution of TDEs.
  For the  first low-redshift detection
  of this class of objects $-$  a jetted TDE $-$
  the fact that  $z_{\rm det} < z_{\rm max}$
  is most likely an effect of small
number statistics (i.e., $N=1$).
  Future observations will shed light on this point.
   (A second
    jetted TDE
     candidate 
    has recently  been
     found 
    with $z=1.2$ [Cenko et al. 2011]). 
}
 (Burrows et al. 2011),
  which encompasses a comoving volume
 $\sim$90 Gpc$^3$.
  Adopting a local galaxy density $0.01$ Mpc$^{-3}$,
  this gives $\sim$10$^9$ galaxies in this volume.
   Taking a 
  nominal 
   tidal disruption rate per galaxy of $3\times10^{-5}$
yr$^{-1}$ (van Velzen et al. 2010)
   and $\xi\simeq0.1$, which lowers the 
  nominal 
  rate 
    by a factor $\sim$$\xi^2$
  to  $3\times10^{-7}$
yr$^{-1}$ galaxy$^{-1}$,
   we obtain a net  rate
   $\sim$300 yr$^{-1}$.
    Adopting 
   a  beaming factor $\sim$10$^{-2}$
reduces the rate to  $\sim$3 yr$^{-1}$,
   and taking a  {\it Swift}/BAT
   sky coverage of   $\sim$10\%
   gives a further reduction
   to  $\sim$ 0.3 yr$^{-1}$.
  Note that the standard TDE rate predicts $\sim$30 BAT events
 per  year (for a beaming factor $\sim$10$^{-2}$),
 which can be strongly excluded.
  This suggests that only a fraction of TDEs can
 create powerful relativistic jets
  ($\sim$10\%; Burrows et al. 2011) and
        reinforces
   the notion of a deep plunging
  orbit as a way of reducing the overall TDE rate.

{\bf {\it Beppo-Sax} rate:}
{\it Beppo-Sax} had a FoV $\sim$0.12 sr (FWHM),
   a sensitivity of 2mCrab in 100 ks,
 a bandpass of $2-10$ keV, and a lifetime 6 yr.
Based on the steady state observed in XRT,
 Sw 1644+57
  could have been
detected up to a redshift $z=1.1$ by the
 {\it Beppo-Sax}/WFCs.
Considering the much smaller 
  FoV (0.12 sr vs. 1 sr of BAT), the rate of 
events
is only 0.2 in 6 yr of the mission, 
 so the fact that Beppo-SAX did
not see an event like Sw 1644+57 
   is consistent with the BAT rate.
By using the brightest peaks,
   the event could have been
observed up to higher redshift ($z\simeq2.5$),
  but  it would have been
impossible to identify it as
   a      TDE because  the long lasting soft
emission would have been under detection threshold.
In any event, even at $z=2$,
    the increase in volume
does not compensate the difference in FoV.

{\bf 
Wide-field X-ray monitor (WFXM) rate:}
  Future mission concepts 
   with sensitive wide-field
   X-ray monitors envision $\sim$1 sr sky coverage
at $\sim$mCrab sensitivity.
 For specificity, we adopt
   a FoV of 0.6 sr,
  a sensitivity of 1.5 mCrab in 100 s,
and a bandpass $0.3-6$ keV.
 Using the peak flux $\sim$10$^{-8}$  erg cm$^{-2}$ s$^{-1}$,
the event could be detected out to $z\simeq4$.
  Based on a  BAT detection rate of $1/7$ yr$^{-1}$,
the  WFXM  rate would be  $\sim$$1-2$ yr$^{-1}$.
  Currently operating X-ray/ASM instruments
  such as the {\it RXTE}/ASM  which  has $\sim$50 mCrab sensitivity
  in $\sim$$6\times10^5$ s 
(for blind searches)
in the bandpass $2-12$ keV
   would yield much lower rates, therefore their
nondetections of these type of events
    are not unexpected.

\subsection{Falsifiability of the TDE interpretation}

In the picture of accretion by shredded stellar remnants,
$L\propto t^{-5/3}$ (Rees 1988), whereas  
   in the accretion disk description
$L\propto t^{-4/3}$ for a super-Eddington disk
  with $h/r\simeq 1$ (CG09).  
The X-ray flux to date
has shown a definite decrease, 
   accompanied by 
  large fluctuations. 
    There are indications from comparing the 
current decay to the early light curve
    that a transition may have occurred around
  $t\simeq 6$ d, which we tentatively
   interpret as the transition to disk decay.
   Therefore we should now (as of this writing, at
  $t\simeq T0+100$ d) be in the regime for
  which $d\log F/d\log t \simeq -4/3$.
 The data to date do not definitely substantiate
  this prediction, but are generally consistent.
   There are also large variations in the flux which may
  reflect jet instabilities (McKinney \& Blandford 2009)
     rather than variations
  in accretion rate onto the central engine.
   The present level of X-ray flux is about  a factor $10^2$
   above the {\it Swift}/XRT detection limit, therefore one could envision
  a more definitive decay law  becoming manifest eventually. 
  If no such law
emerges or if a different law emerges, 
   that would argue against the TDE interpretation.
  If on the other hand 
 the flux drops suddenly to an  unobservable level, 
  that may be consistent with a sudden change
   in the properties of the jet 
  as the accretion rate  drops below Eddington.
   Such a change is seen in the radio properties
  of X-ray binaries as they undergo state
  changes, e.g., from the high/thermal state to the low/hard state.
  For completeness, we note that
    a third option could be  that nonlinear outcome
  of the TDE is completely different from what we describe,
   in which case the nondetection of         the               $t^{-4/3}$ law
      would not be an argument against the TDE.
    In addition,  even though the X-ray flux is currently
dominated by the jet,  it may be feasible to observe
a thermal soft X-ray component from the disk.
   Using Equation (19), our disk model yields at $t=6$ d
   a temperature at the ISCO (innermost stable circular
  orbit)  $6G M_{\rm MBH}/c^2$ of $\sim0.09$ keV $m_{b7}^{-1/2}$,
  which would produce a thermal component peaking
at $\sim$0.2 keV and contributing $\sim$$10^{-2}$
  in flux
   to the overall spectrum.
 According to our estimates a
  $\sim$0.1 keV thermal component
 contributing  only $\sim$1\%
     to the total observed emission 
 could not be detected in the XRT spectra,
    which are dominated by the jet
 emission (Burrows et al. 2011).

\smallskip
\smallskip

\section{Discussion and Conclusion}
 
\smallskip
\smallskip

We have presented simple scalings laws for
 the 
   post-TDE evolution. 
  Three independent lines of reasoning
    point to a deeply plunging orbit
   $R_P \ll R_T$ if the TDE model is to apply
to Sw 1644+57.

\smallskip

(1) 
 An application of the star fallback rate $t^{-5/3}$
to the early Sw 1644+57 light curve (as in Levan et al. 2011b) 
  gives $t_{\rm fallback} \la 1$ d. This implies that the orbit of the 
  disrupted star must have been deeply plunging,
  $\xi \simeq 0.1$.
   If  $R_{\rm P}\approx GM_{\rm MBH}/c^2 $   
  the commonly used Newtonian estimate
   for $t_{\rm fallback}$ would need correcting (Ayal et al. 2000)
   because a standard Schwarzschild BH could not accommodate
  $\xi < 0.2$, for $m_{b7}=1$. 
      This implies a maximal Kerr BH.
  Two first order effects would be 
    Lense-Thirring precession
   and the general relativistic advance of
  the lines of apsides  of ejecta trajectories 
  (Bardeen, Press, \& Teukolsky 1972, 
   Bardeen \& Petterson 1975).
 For  $m_{b7}=0.1$ the minimum allowable $\xi$ value 
  (from Equation [3]) would 
   decrease 
   to $\sim$0.04 for a Schwarzschild BH
   and
      $\sim$0.02 for a Kerr BH,
   lessening the stringency imposed by a $\xi\simeq 0.1$ constraint. 

\smallskip

(2)
   The long term X-ray  luminosity
  is  $\sim$10$^2-10^3$ times greater than
     expected for Eddington accretion.
   This points strongly to beaming, as has been 
    inferred for GRBs.
   In addition, in the long term light curve
   there appears to be a change in the
 decay properties at $t\simeq 6$ d
   which we interpret as the onset of disk-dominated accretion.
   By comparing our
  theoretical
  decay rates for stellar debris
  fallback and accretion in a freely expanding
  super-Eddington disk, 
  we find that a small value $\xi=R_P/R_T$
     is required, $0.1-0.2$,
  for $M_{\rm MBH} \simeq 10^6 - 10^7 \msun$.
  The observed X-ray luminosity
   at $\sim$2.5d
    $L_X/L_{\rm Eddington} \sim$10$^2$ ($10^3$) for $m_{b7}=1$ (0.1)
           would imply a beaming factor 
  $\epsilon_b\simeq10^{-2}$ (for $f=\epsilon_j=0.1$),
     similar to blazars 
 (Sikora et al. 2005, B\"ottcher et al. 2007). 
%
    The sudden introduction of $\sim$0.5$\msun$ of gas
   inside the ergosphere of a Kerr BH would launch 
powerful jets within a few local orbital time scales
  (McKinney 2005).
  Figure 1  of McKinney shows the steep dependence
  of jet efficiency on BH spin to mass ratio $J/M$,
 and in particular a steep upturn close
  to $J/M \simeq1$.
    In fact it may be that 
    our putative constraint on $R_P$,
   namely $R_P\approx GM_{\rm MBH}/c^2$
 (which requires $J/M \simeq1$),
  is a necessary condition
   for this rare event.  

\smallskip

(3) The standard TDE rate $3\times 10^{-5}$  yr$^{-1}$  galaxy$^{-1}$
    implies a {\it Swift}/BAT detection 
  rate $\sim$30 yr$^{-1}$, for a beaming factor $\sim$10$^{-2}$,
   which can be strongly excluded.
   A deeply plunging orbit, $\xi\simeq0.1$, lowers the 
TDE rate by a factor $\sim$10$^{2}$, giving a rate
  in line with the BAT detection rate of $\sim$1/7 yr$^{-1}$.
  Our revised
    rate $\sim$$3\times10^{-7}$ yr$^{-1}$
   galaxy$^{-1}$
    may need adjusting to 
correct for relativistic effects. 
    Ayal et al. (2000)
compare Newtonian and relativistic TDE methods
and find slight differences in $R_P$ between the two 
 sets of calculations, for the same initial
   trajectory. 
    They also find that $\sim$75\% of
the star becomes unbound, rather than $\sim50$\%.
  However, this depends strongly on the treatment of
   cooling, which was not modeled in their SPH (smoothed
  particle hydrodynamics) simulations.
   Another potential source of concern in
our estimates is our  use of the SS73-like algebraic scalings
 for the accretion disk.
   The SS73 methodology was expanded by Novikov \& Thorne
  (1973)
 to include general-relativistic corrections
   including the full Kerr metric. Their
   multiplicative correction factors are based
on Taylor expansions of the form $1 + ax + bx^2 + cx^3 + ...
  + a^{'}y + b^{'}y^2 + c^{'}y^3 ...$,
where $x=GM_{\rm MBH}/(Rc^2)$, $y=J/M$, 
  and the factors $a$, $b$, $c$, ...,
              $a^{'}$, $b^{'}$, $c^{'}$, ... 
          are of order unity.
  These extra  multiplicative terms would therefore modify our simple
  estimates by  factors of order unity,
  even at small radii.

It is worth noting that nearly all 
   previous theoretical
    studies of tidal disruption
  have, for simplicity,
    considered orbits for which the 
 periastron  radius $R_P$
   equals or is 
slightly interior to $R_T$ (Rees 1988, 
     Evans \& Kochanek 1989,
    CLG90, 
   Ulmer 1999,
    SQ09). Few works have considered
 deeply plunging disruptions (e.g., Guillochon et al. 2009).
It appears 
   Sw 1644+57 may require $R_P \ll R_T$,
   and would  be different in at least two important ways
  from the standard TDE.
  (1) First, the rates would
  be lower. 
   The rate for deeply plunging orbits
      would be less than that for  traditional
  TDEs in which $R_P \simeq R_T$ by   
   $\sim (R_P/R_T)^2$, thus giving $\sim$$3\times 10^{-7}$ yr$^{-1}$
   galaxy$^{-1}$, if the more common normal,
   i.e., ``total'' TDE rate integrated over all $R_P$ values
   is 
    $\sim$$3\times 10^{-5}$ yr$^{-1}$
   galaxy$^{-1}$. 
  (2) Second, to observe the event we need to be 
   situated within the beam, which would only
cover $\sim$$10^{-2}$ of the sky.
   Our estimate of the beam size is based on 
the ratio of observed  X-ray luminosity to
   expected mass-energy accretion luminosity.
  There have been no detailed theoretical studies
  investigating the scenario we outline in this work.

  For an orbit in which $R_P/R_T = 0.15$, for example,
the tidal force acting on the star at periastron 
would be $\sim$(0.15)$^{-3} \simeq 300$ times that
  minimally needed to 
  unbind the star.  The subsequent evolution
might be more properly called ``obliteration''
rather than ``disruption'', hence we propose
the acronym TOE (tidal obliteration event).
  From SQ09 and earlier studies,
   in the moving frame of the center-of-mass
  of the star, 
   the acceleration of the outer layers
    due to the MBH tidal force 
 $\sim$$(GM_{\rm MBH}/R_P^2)(R_*/R_P)$
 acting over a dynamical time $t_P \simeq 
 (GM_{\rm MBH}/R_P^3)^{-1/2}$
  results   in velocity perturbations
    to the stellar envelope $\Delta v_P \simeq v_* (R_T/R_P)^{3/2}$,
where $v_*$ is the escape speed from the stellar surface.
   For $(R_P/R_T)\simeq 0.15$, the velocity 
  perturbation exceeds the stellar escape speed
  by a factor $\sim$20.
   The star would likely be completely shredded 
in $\sim$10$^2$ s.  The half of the star
  remnants with orbits on 
inward  trajectories would immediately establish 
   a  
   super-Eddington
  disk within the ergosphere of the MBH. 
%

The discovery of a new class of   objects raises
  several interesting questions.
   For instance, why should a sudden accretion event
   onto a $\sim$$10^6-10^7\msun$  BH resemble in any way
   a GRB?  If there were
  an accretion event in which $\sim$$0.5\msun$
 were suddenly
   introduced near the event horizon,
       the time scales would 
be consistent with what is observed.
Within the first five days
the light curve of Sw 1644+57 showed flares with rise and decay
times of $\sim$$100-200$ s.
    It is unclear whether these variations stem from 
  jet instabilities or are associated with disk time scales.
An orbital time
scale at the tidal radius 
  $2\pi/\Omega(R_T)= 1.5$ hr $r_{13}^{3/2} m_{b7}^{-1/2}$
  (0.5 hr for $m_{b7}=0.1$).
    An after-the-fact examination of prior BAT data 
   revealed a slow brightening over several days, significant
  at the  $\sim$$2-2.5\sigma$ level.
 The source was already bright in X-rays
  by the time XRT began observations.
  It appears we require  a deeply plunging orbit,
   $\xi \simeq 0.1$, and the orbital time $2\pi r/v_\phi$
     at the corresponding $R_P$ is $\sim$300 s for $m_{b7}=1$
    ($\sim$100 s for $m_{b7}=0.1$).
  The small periastron radius and
     general relativistic corrections such
  as the advance of the line of apsides
    would
      enormously speed up
  the TDE evolution compared to what is shown
  for instance in Figure 3 of SQ09.
   This is evidenced in the apparent $\sim$$1-10$ hr timescale
for the fallback time.
    The process
     responsible for utilizing the accretion
    power in the inner disk to tap 
  the BH spin, presumably
  the 
  magnetorotational instability
  (Balbus \& Hawley 1998)
  operating inside the ergosphere
(McKinney \& Narayan 2007ab),
   would proceed with an initial growth rate
 given roughly by the local orbital frequency.
  Given a small seed field within the gas,
it is likely that a large number $\ga$30
    of $e-$folding times
  are required  before the shear amplification
 of the local magnetic field 
    leads to the nonlinear 
   development of a strong jet. 
  Weak precursor events may have contributed
  to low-level BAT activity preceding the main
  Sw 1644+57 trigger.
   Considering the commonalities among jets
    observed to date
in such a wide variety of objects,
  Nature appears to be telling us that the jet physics is
largely independent of the mass of the central black hole.

\smallskip

We thank L. Piro and E. Rossi for useful
   conversations.
This work made use of data supplied by the UK {\it Swift}
   Science Data Centre at the University of Leicester.
ET was supported by an appointment
  to the NASA Postdoctoral Program
  at the Goddard Space Flight Center,
administered by the Oak Ridge Associated Universities
  through a contract with NASA.

\vfil\eject

\end{document}